\documentclass[12pt,a4paper,reqno]{article}
\pdfoutput=1
\usepackage[english]{babel}
\usepackage{amssymb,graphicx}
\usepackage[pdfstartview=FitH]{hyperref}

\renewcommand{\thesubsection}{{\bf\arabic{section}.\arabic{subsection}}}
\makeatletter
\renewcommand{\section}{\@startsection {section}{1}{\z@}%
             {-3.5ex \@plus -1ex \@minus -.2ex}%
             {2.3ex \@plus.2ex}%
             {\normalfont\normalsize\sffamily\bfseries}}
\renewcommand{\subsection}{\@startsection {subsection}{1}{\z@}%
             {-3.5ex \@plus -1ex \@minus -.2ex}%
             {2.3ex \@plus.2ex}%
             {\normalfont\normalsize\sffamily\emph}}

\renewcommand\theequation{\arabic{section}.\arabic{equation}}

\@addtoreset{equation}{section}
\makeatother
 \usepackage[isolatin]{inputenc}
 \usepackage{epsf,amsfonts}
 \usepackage{fontenc,indentfirst, delarray,epstopdf}
 \usepackage{a4,amsmath,amssymb,amsfonts}
 
\begin{document}
 \newcommand{\norm}[1]{\left\lVert#1\right\rVert}
\newcommand{\abs}[1]{\lvert#1\rvert}
\newcommand{\scal}[1]{\langle#1,#1\rangle}
\newcommand{\scl}[2]{\langle#1,#2\rangle}
\newcommand{\suup}[1]{ \underset{#1}{\sup} }
\newcommand{\grad}[1]{\text{grad}\,#1}
\newcommand{\ket}[1]{\lvert#1\rangle}
\newcommand{\bra}[1]{\langle#1\lvert}
\newcommand{\tr}[1]{\text{Tr}(#1)}

\def\re{\text{Re}}
\def\im{\text{Im}}

\def\dm{\lp\begin{array}}
\def\fm{\end{array}\rp}
\def\dbb{\lb\begin{array}}
\def\fbb{\end{array}\rb}
\def\dbn{\left.\begin{array}}
\def\fbn{\end{array}\right.}

\def\beq{\begin{equation}}
\def\eeq{\end{equation}}

\def\ee{{\mathcal E}}
\def\lb{\left[}
\def\rb{\right]}
\def\lp{\left(}
\def\rp{\right)}
\def\alg{alg\`ebre }
\def\nc{non commutative}
\def\gnc{g\' eom\' etrie \nc}
\def\ms{mod\`ele standard}

\def\calg{$C^*$-\alg}
\def\dn{\partial_{\nu}}
\def\ds{\slash\!\!\!\partial}
\def\da{\left[ D,\pi(a) \right]}
\def\df{\left[ D,f \right]}

%********** ensembles *************
\def\m3{M_3 \lp \cc \rp}
\def\m2{M_2 \lp \cc \rp}
\def\mmp{M_p \lp \cc \rp}
\def\mn{M_n \lp \cc \rp}
\def\mnp{M_{np} \lp \cc \rp}
\def\mpn{M_{pn} \lp \cc \rp}
\def\mx{M_k}
\def\mz{M_{k'}}
\def\kk{{\mathbb{K}}}
\def\cc{{\mathbb{C}}}
\def\rr{{\mathbb{R}}}
\def\nn{{\mathbb{N}}}
\def\zz{{\mathbb{Z}}}
\def\ii{{\mathbb{I}}}
\def\iii{{\mathcal{I}}}
\def\aa{{\mathcal A}}
\def\bb{{\mathcal B}}
\def\ll{{\mathcal L}}
\def\bbb{{\aa_I}}
\def\dd{{\mathcal D}}
\def\hh{{\mathcal H}}
\def\jj{{\mathcal J}}
\def\oo{{\mathcal O}}
\def\ss{{\mathcal S}}
\def\mm{{\mathcal M}}
\def\pp{{\mathcal P}}
\def\ccc{{\mathcal C}}
\def\hhh{{\mathbb H}}
\def\jj{{\mathcal J}}
\def\gg{{\mathcal G}}
\def\xx{{\mathcal X}}
\def\ginf{\Gamma^\infty}

\def\cinf{C^{\infty}\!\!\lp\mm\rp}
\def \difm{\text{Diff}({\mathcal{M}})}
\def\cm{C^{\infty}(\mm)}
\def\spin{\text{Spin}}
\def\del{\triangledown}
\def\L2{L_2(\mm)}
\def\LS{L_2(\mm , S)}

\def\cnp{\cc^{np}}
\def\cpn{\cc^{pn}}
\def\cn{\cc^{n}}
\def\cp{\cc^{p}}

%******* vecteurs et etats **************
\def\ot{\otimes}
\def\pof{\psi\otimes\xi}
\def\fop{\xi\otimes\psi}

\def\xox{{\xi}_x}
\def\yox{{\xi}_y}
\def\xoz{{\zeta}_x}
\def\yoz{{\zeta}_y}

\def\omx{\omega_x}
\def\omy{\omega_y}

\def\ox{\omega_{\xi}}
\def\oz{\omega_{\zeta}}
\def\xo0{\omega^0_x}
\def\yo0{\omega^0_y}
\def\tox{\tilde{\omega_{\xi}}}
\def\toz{\tilde{\omega_{\zeta}}}
\def\oe{\omega_{E}}
\def\oi{\omega_{I}}
\def\oei{\omega_{E}\ot\omega_{I}}
\def\oeip{\omega_{E}\ot\omega_{I}'}
\def\oepi{\omega_{E}'\ot\omega_I }
\def\oepip{ {\omega_E}' \ot {\omega_I}'}
\def\xoi{\omega_x\ot\omega_i}
\def\xoip{\omega_x\ot\omega_{i}'}
\def\yoi{\omega_y\ot\omega_i}
\def\yoip{\omega_y\ot\omega_{i}'}

\def\ou{{\omega_{1}}}
\def\od{{\omega_{2}}}
\def\odu{\omega_{21}}
\def\oud{{\omega_{12}}}
\def\opud{{{\omega'}_{12}}}
\def\opdu{{{\omega'}_{21}}}
\def\ou{{\omega_{1}}}
\def\ouu{{\omega_{11}}}
\def\odd{{\omega_{22}}}
\def\oc{{\omega_{c}}}
\def\oeu{\omega_E \ot \ou}
\def\oed{\omega_E \ot \od}
\def\oeo{\omega_E \ot \omega}

\def\oxe{\omega_{k_e}}
\def\oze{\omega_{{k'}_e}}
\def\oue{\omega_{1}\circ\alpha_e}
\def\ode{\omega_{2}\circ\alpha_e}
\def\o0{\omega_0}
\def\rx{\rho_\xi}
\def\rz{\rho_\zeta}
\def\sx{s_\xi}
\def\sz{s_\zeta}

\def\xox{\xi_x}
\def\yox{\xi_y}
\def\rk{\rho_k}
\def\rkp{\rho_{k'}}
\def\ux{{u_k}}
\def\uz{{u_{k'}}}
\def\xoz{x_{{k'}}}
\def\yoz{y_{{k'}}}
\def\xo0{x_\omega^0}
\def\yo0{y_\omega^0}
\def\tox{\tilde{\omega_{k}}}
\def\toz{\tilde{\omega_{{k'}}}}
\def\nx{{n_k}}
\def\nz{{n_{k'}}}

\def\ae{\alpha_e}
\def\aea{\alpha_e(\aa)}
\def\fu{\alpha_u}
\def\fue{\alpha_{u^*}}
\def\tu{\tau_1}
\def\td{\tau_2}
\def\futu{\fu(\tu)}
\def\futd{\fu(\td)}
\def\fuetu{\fue(\tu)}
\def\fuetd{\fue(\td)}
\def\tue{\tu\circ\ae}
\def\tde{\td\circ\ae}

\def\te{\tau_E}
\def\ti{\tau_I}
\def\tei{\tau_E\ot\tau_I}
\def\po{\pi_\omega}
\def\pto{\tilde{\pi}_\omega}

\def\auta{\text{Aut}(\aa)}
\def\ina{\text{In(\aa)}}
\def\outa{\text{Out}(\aa)}\def\auta{\text{Aut}(\aa)}
\def\inmn{\text{In}(\mn)}
\def\outmn{\text{Out}(\mn)}

\def\cl{\text{Cl}}
\def\ccl{\cc\text{l}}

 % \markboth{Pierre Martinetti}
%  {Line element in quantum gravity}

% %%%%%%%%%%%%%%%%%%%%% Publisher's Area please ignore %%%%%%%%%%%%%%%
% %
% %\catchline{}{}{}{}{}
% %
% %%%%%%%%%%%%%%%%%%%%%%%%%%%%%%%%%%%%%%%%%%%%%%%%%%%%%%%%%%%%%%%%%%%%

\title{Line element in quantum gravity: the examples of DSR and noncommutative geometry}

\author{Pierre Martinetti{\footnote{martinetti@physik.uni-goe.de}}\\
\\
%}
%
%\address{
{ Institut f\"ur Theoretische Physik,}\\
{  Georg-August Universit\"at, Göttingen}}
\maketitle
% % \begin{history}
% % \received{Day Month Year}
% % \revised{Day Month Year}
% % \end{history}
\begin{abstract}
We question the notion of line element in  some quantum 
spaces that are expected to play a role in quantum
 gravity, namely non-commutative deformations of Minkowski spaces.
We recall how the implementation of the Leibniz rule forbids to see 
some of the infinitesimal deformed Poincaré transformations as good candidates
for Noether symmetries. Then we recall the more fundamental view
on the line element proposed in noncommutative geometry, and re-interprete at this
light some previous results on Connes' distance formula.

%\keywords{Keyword1; keyword2; keyword3.}
\end{abstract}

%\ccode{PACS numbers: 11.25.Hf, 123.1K}

\section{Introduction}	
 In this talk we would like to illustrate how the line element survives
and/or is modified in quantum gravity. This is a relevant question in order to understand how the notion of distance may survive or be abandoned
along the process of quantizing gravity. Here ``quantum gravity'' has to be intended as an area of research rather than 
one precise physical theory since, as everybody know, there is still no well established theory 
describing a quantized gravitational field. Many, if not all, the approaches to quantum gravity rely
on the idea that geometry at small scale, e.g. the Planck scale, should be modified. In particular it is largely admitted
that the description of space-time as a manifold is no longer viable when both quantum and gravitational
effects are taken into account. As a consequence one may either renounce geometry as a useful tool to do physics (as Weinberg proposed 
in \cite[chap. 6.9]{weinberg}), or on the contrary take quantum gravity requirements as hints indicating where to push geometry in order to get fruitful mathematical developments. 
 Among these requirements, one most
 widespread in the scientific community (as well as in a larger audience) is the idea of extra-dimensions, intensively popularized by string 
theory. Another requirement maybe less widespread, at least in the grand public, is noncommutativity. Let us see what happens to the line element in both contexts.

  Adding a finite number of dimensions to four dimensional Minkowski space-time is not creating any difficulties in defining the line element according to the usual 
formula
\begin{equation}
  \label{eq:2}
  ds^2 = g_{\mu\nu} dx^\mu dx^\nu
\end{equation}
as soon as one is able to provide a value to the extra-coefficients of the metric $g_{\mu\nu}$ for $\mu, \nu > 4$. This may be a complicated task, depending 
on the physics governing the extra-dimensions, but there is in principle 
 no obstruction against the use of (\ref{eq:2}) as length element in the presence of extra-dimensions.
As a consequence the notion of distance, obtained by integrating $ds$ along a minimal geodesic, is also a priori meaningful. 

The situation is much more involved regarding noncommutativity since neither the metric tensor nor the $1$-form $d x^\mu$ have an obvious noncommutative translation.
A strategy could be to define a noncommutative differential calculus, and a noncommutative metric tensor, hoping that the combination of the two still makes sense as a noncommutative line element. There is a vast amount of litterature discussing noncommutative differential calculus (e.g. \cite{woronowicz}, \cite{dubviol}, \cite{madore}) and several proposals on how to define a corresponding metric and line element (e.g. \cite{Vassilevich, Ghosh:2007hc, buricmadore}) but, to the knowledge of the author, there are very few proposals on how to extract from those an effective notion of distance. One of these proposals is Connes' definition\cite{CL92} of the the line element in \emph{noncommutative geometry}\cite{Con94}
 in an operatorial way, namely as the inverse $D^{-1}$ of a generalized Dirac operator, yielding the distance formula (\ref{eq:11}).

 In the following we illustrate this variety of points of views by two examples. One is based on a pedestrian approach to Noether theorem on \emph{noncommutative spaces} (i.e. non-commutative deformations of Minkowski space-time) and show
how imposing the Leibniz rule forces to restrict the set of acceptable line elements. 
The second is the already mentioned definition of the line element  in \emph{noncommutative geometry}, $ds = D^{-1}$. Although more abstract at first sight, this definition actually preserves the effectiveness of the line element, in the sense that it allows to compute \emph{ explicitly} distances in a noncommutative framework. We provide various examples of such computations. 

\section{Line element in  noncommutative spaces: the example of DSR}
\subsection{Deformed Minkowski spaces and deformed Poincaré algebras}
Since the very beginning of quantum mechanics, one knows that phase space cannot be described as a ``space'' in the classical sense since the coordinates $x, p$ generates
a non-commutative algebra. Various works (e.g. \cite{dfr}) further suggest that space-time itself should be viewed as non-commutative at small scale. For a systematic study on which noncommutative algebras could be considered as viable non-commutative coordinates, one should refer to \cite{dubviol2}. Here we limit ourselves to two kinds of noncommutative 
deformations of Minkowski space-time ${\mathcal M}$ that are rather popular in physics. One is the $\theta$ deformation, that appears for instance in string theory,
  \begin{equation}
[x_\mu,x_\nu] = i\theta_{\mu\nu}
\label{eq:40}
\end{equation}
where $\theta$ is a antisymmetric matrix (in the example below we restrict to an observer independent constant $\theta$). Other is a Lie-algebraic deformation
 \begin{equation}
[x_0,x_i] = \frac{i}{\kappa} x_i, \quad [x_i, x_j] = 0
\label{eq:4}
\end{equation}
where $\kappa\in\rr^*$. Such $\kappa$-Minkowski spaces are of particular interest in quantum gravity for they may provide a concrete implementation of the idea of Doubly (or deformed) Special Relativity\cite{Amelino-Camelia:2002jw}. DSR consists in building a theory in which, besides the speed of light $c$, there is another reference-frame independent quantity, typically a length $\lambda$ further identified to the Planck length $l_p = \sqrt{\frac{\hbar G}{c^3}}\simeq 1.62\, 10^{-35}m$. Following Snyder's idea\cite{Snyder:1947ix}, $\kappa$-Minkowski is a good candidate for DSR by simply taking as a deformation parameter $\kappa = \lambda^{-1}$. More recently $\kappa$-Minkowski spaces also appeared as effective theories describing the coupling of matter in certain spin-foam models\cite{freidel}. In the following we consider both $\theta$ and $\kappa$ deformations for our considerations on the line element are similar in both cases.  

The  symmetries of a deformed Minkowski space-time are described by a deformation of classical Poincaré symetries.
Namely one considers an algebra of generators $\left\{ P_\mu,  M_{\mu\nu}\right\}$ \-- $P_\mu$, $\mu\in[0,4]$, are the generators of translations, $M_{ij}$, $i,j \in [1,3]$, are those of rotations,
$M_{0j}$ are boosts \-- that acts on the noncommutative coordinates like ordinary
Poincaré generators,
\begin{equation}
\label{actionclass}
  P_\mu x_\nu = i\eta_{\mu\nu},\quad  M_{\mu\nu} x_\alpha = ix_{[\mu}\eta_{\nu]\alpha}
\end{equation}
where the brackets denote the anti-symmetrization on the indices and $\eta$ is the Minkowski metric.
The deformation appears when one considers a product of coordinates. Indeed assuming that any generator $N$ satisfy 
 the Leibniz rule, 
$N(x_\mu x_\nu) = (N x_\mu)x_\nu + x_\mu N x_\nu$, is not compatible with
 (\ref{eq:40}) or (\ref{eq:4}). For instance
\begin{equation}
  \label{coprot}
  M_{\mu\nu} ([x_\alpha,x_\beta]) = [M_{\mu\nu}x_\alpha , x_\beta] + [x_\alpha, M_{\mu\nu}x_\beta]
 %=  [ix_{[\mu}\eta_{\nu]\alpha} , x_\beta] + [x_\alpha, i x_{[\mu} \eta_{\nu]\beta]}]
= \theta_{\beta[\mu}\eta_{\nu]\alpha} - \theta_{\alpha [\mu}\eta_{\nu]\beta]}
\end{equation}
is non zero (just take $\mu=\alpha=1$, $\nu= 2$, $\beta =3$) while $M_{\mu\nu}(\theta_{\alpha\beta}) = 0$.
Hence one has to deform the action of generators on a product of coordinates. Such a deformation is best captured
in the formalism of Hopf algebra,
by defining the action on a product of coordinates in term of a coproduct $\Delta$,
\begin{equation}
  \label{eq:6}
  N(x_\mu x_\nu) = \Delta N(x_\mu \otimes x_\nu)\quad\text{ where }\quad
  \Delta N = \sum_i  N_{1i} x^\mu \otimes  N_{2i}\, x^\nu
\end{equation}
with $N_{ji}$ operators (typically generators or exponentials of generators) acting on the deformed-Minkowski 
space-time via (\ref{actionclass}).
To say it shortly a deformation of Poincaré algebra amounts to a prescription of coproducts for the generators.
 For instance to a $\theta$-deformation corresponds the $\theta$-Poincaré algebra characterized by
\begin{equation}
\label{coproduct}
 \Delta P_\mu \! =\! P_\mu \otimes 1 + 1\otimes P_\mu,\;
 \Delta M_{\mu\nu}\! =\! M_{\mu\nu} \otimes 1 + 1\otimes M_{\mu\nu}\! -\! \frac 12\theta^{\alpha\beta}\left[ 
\eta_{\alpha [\mu} P_{\nu]}\!
\otimes \!P_\beta + P_\alpha\! \otimes\! \eta_{\beta[\mu} P_{\nu]}\right].
\end{equation}
One easily checks that (\ref{eq:40}) is preserved under the action of such generators. This is obvious for translations since by (\ref{coproduct}) the latter are not deformed and the commutator (\ref{eq:4}) behaves with respect to translation as a classical (i.e. vanishing) commutator. For rotations and boosts one checks that the non-trivial part of the coproduct exactly cancels (\ref{coprot}). Note that this deformation can be understood as a twist of ordinary Poincaré algebra.
For the Lie algebraic deformation, a deformed $\kappa$-Poincaré algebra preserving (\ref{eq:4})  is
\begin{equation}\label{coproduct2} \Delta P_i = P_i \otimes 1 + e^{-\frac{P_0}\kappa}\ot P_i, \quad 
 \Delta N_i= N_i \otimes 1 + e^{-\frac{P_0}\kappa}\ot N_i +  \frac{\epsilon_{ijk}}\kappa P_j \otimes M_k,
\end{equation}
where $M_k\doteq \epsilon_{kmn} M_{mn}$ generates rotation around the $k$-axis and $N_j \doteq M_{0j}$. Time-translation and rotations are not deformed.

Scalar fields  $\phi$ (or more generally functions) are defined on noncommutative Minkowski spaces through the mapping of functions on ${\mathcal M}$ via a Weyl map $\Omega$\cite{madorewess},
 \begin{equation}
   \label{eq:1}
\phi (x) \doteq \int \tilde{\phi}(p)\, \Omega(e^{ipx}) \, d^4p   
 \end{equation}
where $\tilde\phi$ is the Fourier transform of a function on $\mm$ and  $px$ denote the product of $4$-vector. There exist different Weyl maps, corresponding to different ordering in the exponential.
%, like $\Omega_R (e^{ipx}) = e^{i p.x - p_0x_0}$ or $\Omega_S (e^{ipx}) = e^{-i\frac{p_0x_0}2} e^{i p.x} 
%e^{\frac{ip_0x_0}2}$ 
Here we chose the Weyl ordering, $\Omega(e^{ipx}) = e^{-i p_0x_0} e^{i p.x}$ with $p.x$ the product of 
spatial $3$-vectors. This turns out to be compatible with the coproducts above, in the sense that $\Omega (N (xy)) = N\Omega(xy)$ for any generators $N$ (viewed as elements of the resp. classical resp. deformed Poincaré algebra in the resp. l.h.s. resp. r.h.s. of the equation).  Other Weyl maps yield other coproducts, but this apparent ordering-dependence simply corresponds to a change of bases in the algebra. 
\subsection{Noether analysis}

Let us begin by recalling basics of Noether analysis in the classical space $\mm$.
Consider the variation of the action $\iii =\int d^4x \,\ll(\phi(x))$ for some lagragian density $\ll$ on $\mm$ under a combined transformation of the both the coordinates
and the field,
\begin{equation}
\label{fields0}
 x\mapsto x' ,\quad \phi \mapsto  \phi'.
\end{equation}
Specifically let us choose a change of coordinates coming from the action on $\mm$ of some  
 $\Lambda = e^{ta}$ in the Poincaré group $G$,
\begin{equation}
  \label{eq:9}
 x\mapsto \Lambda x \doteq  \sigma(t,x) = e^{ta}x 
\end{equation} 
with $t\in\rr, a \in\frak g$ and $\sigma$ the flow generated by $a$ on $\mm$. 
 For any smooth function $f$ on $\mm$ the action of $G$ yields an infinitesimal variation
\begin{equation}
  \label{eq:10}
  df \doteq f(\Lambda x) - f(x) =   \epsilon V[f]_x \quad\quad\quad t=\epsilon < \!\!< 1
\end{equation}
%where $\epsilon^\mu \doteq \epsilon V^\mu$ with $V^\mu$ the components of $V$. 
where $V$ is the vector tangent to the flow $\sigma$. Combining (\ref{eq:9}) with the scalar action of $G$ on a field, that is to say $\phi\mapsto \phi' \doteq  \phi \circ \Lambda^{-1}$, one gets a vanishing global transformation
(\ref{fields0}) since  $\phi'(x') =  \phi\circ\Lambda^{-1}(\Lambda x) = \phi(x)$. Note that (\ref{eq:10}) written for $\Lambda^{-1}$ equivalently defines the scalar action as the transformation $\phi\mapsto \phi + \delta\phi$ where
\begin{equation}
  \label{eq:14}
  \delta\phi =  - d\phi. 
\end{equation}
 As a consequence $\delta\iii =0$ and Noether 
theorem consists in writing this vanishing variation  as a $4$-divergence
\begin{equation}
  \label{eq:3}
\delta\iii = \epsilon^A \int d^4 x P_\mu J^\mu_A \phi(x) 
\end{equation}
in order to identify Noether courants $J^\mu_A$, whose integration yields Noether charges. To do so one notices
that at first order,
\begin{align}
  \label{eq:5}
  \hspace{-.28truecm}&\delta\iii = \int d^4x\; \delta \ll(x)  + d \ll_{\lvert x} \text{ with}\\
  \label{eq:8}
 \delta \ll (x) \doteq \ll(\phi'(x)) - \ll(\phi(x))\;, &\quad\quad d\ll (x) \doteq \ll(\phi(x')) - \ll(\phi(x)) = \epsilon^\mu(x)\partial_\mu \ll_{\lvert x}
\end{align}
 where in the last equation $\ll$ is viewed as the function $\ll\circ \phi$ of the variable $x$ and
 \begin{equation}
\epsilon^\mu = \epsilon V^\mu
\label{eq:20}
\end{equation}
with $V^\mu$ the component of $V$ in (\ref{eq:10}). 
Then the equation of motion allow to write $\delta \ll$ as a $4$-divergence, hence
(\ref{eq:3}). 

On a deformed noncommutative Minkowski spacetime the strategy is the similar. The infinitesimal action of the deformed Poincaré algebra on $\phi$ makes sense through the Weyl map (\ref{eq:1}), namely 
$N (\phi(x)) = \int \tilde{\phi}(p) N\Omega(e^{ipx})d^4p$. Instead of (\ref{eq:10}) one defines 
  \begin{equation}
    \label{eq:12}
    d\phi = \epsilon^A N_A \phi
  \end{equation}
where $\epsilon^A$ are some infinitesimal coefficients and $N_A$ are generators of the deformed Poincaré algebra. Then by (\ref{eq:14}) one
\emph{defines} the scalar action of $\epsilon^A N_A$ as the map
\begin{equation}
  \label{eq:13}
  \phi \mapsto \phi' = \phi + \delta\phi \text{ with } \delta\phi = -d\phi.
\end{equation}
The point is then to find an action which remains invariant under this scalar action combined with $x\mapsto \epsilon^A N_A x$, then use the equation of motion together with the commutation rules to put the coefficient $\epsilon^A$ on one-side of the integral, so that to obtain an expression similar to (\ref{eq:3}). In $\kappa$-Minkowski and for $\epsilon^A N_A$ a translation, the full program has been achieved initially in \cite{alegiac} then generalized to rotations and boosts in \cite{nopure}. Independently Noether charges for $\kappa$-Minkowski have also been worked out in \cite{freidkowal}. Differences between the two approaches are discussed in \cite{kappa5d}. For  $\theta$-Minkowski Noether charges have been 
computed in \cite{thetan} (see also \cite{aschieri}). In any case, as recalled in 
\cite{giap}, there are some restrictions on the allowed $\epsilon^A N_A$, discussed in the following section.

\subsection{Line element and the no-pure boost principle}   
Consider the Lagrangian  $\ll = \phi\square\phi$ with $\square= P^\mu P_\mu$, together with the scalar action of some infinitesimal transformation $\epsilon^A N_A$. By (\ref{eq:8}) and (\ref{eq:13})
at first order
\begin{align}
\delta\ll = \ll (\phi') - \ll (\phi)  =  \delta\phi \square  \phi +   \phi \square \delta \phi
 =  - (d\phi \square  \phi +    \phi \square d\phi),
\end{align}
so that for the action to be conserved, that is $\delta\ll = -d\ll$, it is necessary that $d$ satisfies the Leibniz rule. But in case $d$ is a single generator $N$, the Leibniz rule is not satisfied unless $N$ has a trivial coproduct. Which means for instance that rotations on $\theta$-Minkowski and boosts on 
$\kappa$-Minkowski are not good candidates for a Noether symmetries. In \cite{nopure} and \cite{thetan} we found that the linear combinations $d = \epsilon^A N_A = \epsilon^\mu P_\mu + \omega^{\mu\nu} M_{\mu\nu}$ that satisfy the Leibniz rule are characterized by simple commutation rules for the infinitesimal coefficients . Namely for $\theta$-Minkowski one gets 
\begin{equation}
  \label{eq:7}
\left[x^\alpha,\omega^{\mu\nu}\right] = 0,\; \left[x^\alpha,\epsilon^\beta\right] = -
\frac{1}{2}\omega^{\mu\nu}\Gamma^{\beta\alpha}_{\mu\nu} \quad \text{ with } \Gamma^{\beta\alpha}_{\mu\nu} \doteq
\theta^\beta_{\left[\right.\mu}\delta_{\nu\left.\right]}^\alpha
-
\theta^\alpha_{\left[\right.\mu}\delta_{\nu\left.\right]}^\beta.
\end{equation}
This means that, as soon as $d$ contains a rotation with component $\omega^{\mu\nu}\neq 0$, the commutator $[x^\alpha, \epsilon^\beta]$ is no zero hence $d$ also contain
a translation. For $\kappa$-Minkowski we found some similar condition that forbid pure boost transformations. 

The nature of the non-commuting infinitesimal parameter $\epsilon^\alpha$ is still not clear. It is tempted to identify it to $1$-form $dx^\alpha$, as done in \cite{freidkowal}. However one has to be careful 
not to identify $d \phi$ to the exterior derivative (and so not to assume that $d^2$ is zero) for, has shown by (\ref{eq:10}), $d$ is in fact the \emph{line element} $dl$ of the curve tangent to the action of $a\in \frak g$ on $\mm$. More exactly with (\ref{eq:20}) the line element
can formally be written as $dl = \norm{V} = \frac{\norm{\epsilon^\mu\partial_\mu}}{\epsilon}$. In the noncommutative case one could write $dl = \frac{\norm{\epsilon^A N_A}}{\epsilon}$ but this expression has no sense as long as the nature of $\epsilon^A$ has not been made clearer.

\section{Line element in noncommutative geometry}
Independently of quantum gravity, mathematics have their own motivation to develop a non-commutative theory of geometry.  
Since at least the work of Gelfand it is known that a space in the traditional sense is a commutative space. More exactly any locally compact topological space $X$ is homeomorphic to the spectrum of the commutative algebra $C_0(X)$ of continuous
functions on $X$ vanishing at infinity, and any commutative  $C^*$-algebra can be viewed as the algebra of functions vanishing at infinity on its spectrum (the spectrum of an algebra $\aa$ being the set of pure states, i.e. positive linear application $\varphi: \aa \mapsto \cc$ with norm $1$, such that $\varphi$ cannot be written as a convex linear
combination of two other such applications $\varphi_1, \varphi_2$). In short any $x$ in $\mm$ is one-to-one associated to $\omega_x$ in the spectrum 
of $\mm$ defined as
\begin{equation}
  \label{eq:16}
  \omega_x(f) = f(x) \quad \forall f\in C_0(\mm).
\end{equation}
 The identification of space with commutative algebra, including not only topology but also differential structure and metric, 
has culminated recently in Connes' reconstruction theorem \cite{Con08} which shows how to characterize a (compact) riemannian spin manifold on purely algebraic settings. The central tool is then not only an algebra $\aa$ but a spectral triple $(\aa, \hh, D)$ including an Hilbert space $\hh$ on which the algebra acts and a selfadjoint operator $D$ acting on $\hh$, which is a generalization of the Dirac operator $\ds = -i\gamma^\mu \nabla_\mu$ of quantum field theory. A spectral triple with commutative $\aa$ describes a Riemannian spin manifold, a spectral triples with a noncommutative $\aa$ describes a geometrical object beyond the scope of Riemannian geometry, 
that one calls a \emph{non-commutative geometry}. Compare to the noncommutative spaces introduced above, noncommutative geometries cover a larger range
of examples since there exist more noncomutative algebras than the deformations (\ref{eq:4}) or (\ref{eq:40}). Connes' theory also provides a 
noncommutative equivalent to all the tools available in 
differential geometry, including the distance (see below). However it is also fair to say that deformations of Minkowski spacetimes allows to treat, at least partially, pseudo-euclidean signature which is, so far, not possible in noncommutative geometry. Some proposal have been made in this direction \cite{kappaspectral}.

As mentioned in the introduction, the line element in noncommutative geometry is defined as the inverse of the Dirac operator. This is an effective definition for it allows, starting from the canonical spectral triple $(\cinf, L_2(M,S),\ds)$ associated to a Riemannian compact manifold $\mm$ (in case $\mm$ is only locally compact, one should consider 
$C_0^\infty(\mm)$ instead) to recover the geodesic distance from purely algebraic datas. Namely one checks that
\begin{equation}
  \label{eq:11}
  d(\omega_x, \omega_y ) = \sup_{a\in\aa} \left\{ \abs{\omega_x(a) - \omega_y(a)},\; \norm{[D,a]}\leq 1\right\}
\end{equation}
coincides with $d_{\text{geo}}(x,y)$ once pure states have been identified with points by (\ref{eq:16}). The advantage of formula (\ref{eq:11}) with respect to the usual definition of the distance as the infimum of the length on all curves between two given points is that (\ref{eq:11}) still makes sense for 
a spectral triple with noncommutative $\aa$ and thus provides noncommutative geometry with a notion of distance between (pure) states of the algebra that i. is coherent with the classical (commutative) case
since then $d=d_{\text{geo}}$, ii. does not rely on some notion ill-defined in a quatum context (like ``shortest path between points'') but only on the spectral properties of the algebra and the Dirac operator. That is why in the following we call $d$ the \emph{spectral distance}.

Noncommutative geometry is flexible enough to describe spaces obtained as the product of a continuum part by a discrete part. Typically this is obtained
 making the product of the 
canonical spectral triple of a manifold with a second spectral triple $(\aa_I, \hh_I, D_I)$ in which $\aa_I$ is a matrix algebra. For the standard model for instance one chooses
$\aa_I = \cc \oplus \hhh \oplus M_3(\cc)$ so that to recover the gauge group of the standard model the unitaries of $\aa_I$ (modulo a lift to the spinors\cite{unimod}).
The picture of space-time that emerges is a then two sheets-model, that is to say two-copies of the manifold $\mm$ indexed by the unique pure state $\omega_0$ of $\cc$ and the unique pure state $\omega_1$ of $\hhh$ (the pure states of $M_3(\cc)$ are all at infinite distance from one another). The distance between two points $x_i, y_j$ ($i,j\in [0, 1]$ index the copies of $\mm$) coincides with the geodesic distance within the manifold $\mm' = \mm \times [0,1]$ with metric\cite{kk}
\begin{equation}
  \label{eq:17}
  \left( \begin{array}{cc} g^{\mu\nu} & 0 \\ 0 & (\abs{1 + h_1}^2 + h_2^2)\; m^2_{\text{top}}\end{array}\right)
\end{equation}
where $g_{\mu\nu}$ is the metric of $\mm$, $\left(\begin{array}{c} h_1\\ h_2\end{array}\right)$ is the Higgs doublet and  $m_{\text{top}}$ is the mass of the quark top.

 The fact that the spectral distance seems to ``see'' between the sheets (although there are no points/pure states between the sheets) can be traced back to the formula given the Dirac operator $D$ in the product of spectral triple, namely
\begin{equation}
  \label{eq:18}
  D = \ds\ot \ii_I + \gamma^5 \ot D_I
\end{equation}
where $\ii_I$ is the identity of $\hh_I$ and $\gamma^5$ the product of the matrices $\gamma$. Taking the square, one obtains
\begin{equation}
  \label{eq:19}
  D^2 = \ds^2 \ot \ii_I + \ii_E \ot D_I^2
\end{equation}
where $\ii_E$ the identity on $L_2(M,S)$. In calculating the distance in the standard model, the first step is to shows that the distance is the same 
as the one in a spectral triple in which $D_I^2 = K^2 \ii_I^2$ where $K= \sqrt{(\abs{1 + h_1}^2 + h_2^2)}\; m_{\text{top}}$ (see \cite{kk} for details), therefore
(\ref{eq:19}) rewrites
\begin{equation}
  \label{eq:21}
  D^2 = (\ds^2 + K^2\ii_E)\ot \ii_I.
\end{equation}
Disregarding the identity and  introducing a ``discrete'' derivative $\partial_0$ such that $\partial_0^2 = \ii_E$ together with a new gamma matrix $\gamma^0 \doteq K\gamma^5$, one can write
\begin{equation}
  \label{eq:22}
  D^2 = (\gamma^A\partial_A)^{2}
\end{equation}
which is the inverse of the line element in $\mm\times$ "a discrete dimension corresponding to the discrete derivative". Hence that the distance is given
by (\ref{eq:17}) is not a surprise. Of course these remarks are not an alternative proof to the exact computation made in \cite{kk}, but it allows to understand the result in a more intuitive way. 

Similar phenomena occurs when one considers the distance encoded within a covariant Dirac operator $D = \gamma^\mu( \partial_\mu + A_\mu)$ on a fiber bundle $P$, with $A_\mu$ the local form of the $U(\aa_I)$-valued $1$-form field associated to the connection on $P$\cite{cc,smoother}. On finds that, although the state space is a circle $S^1$, the distance coincides with the geodesic distance inside the disk\cite{african}. In other terms the spectral distance sees the disk through the circle in the same way as it sees between the two sheets in the standard model.However it seems that this is only valid in low dimension: when the pure states space is a $n$-torus, $n\geq2$, the spectral distance is no longer the euclidean distance inside the torus\cite{cs}. 

\section{Conclusion}
Taking into account noncommutativity of space-time yields different conclusions regarding the line element: on noncommutative space-times it put some restriction on the acceptable line elements, while on noncommutative geometry the notion of line element seems to gain more structure once noncommutativity is allowed.

\end{document}